\documentclass[aps,prl,twocolumn,showkeys,showpacs,nofootinbib]{revtex4}

\usepackage[utf8]{inputenc}
\usepackage{hyperref}
\usepackage{xcolor}
\usepackage{graphicx}
\usepackage{amsmath,amssymb,amsbsy,amstext,amsthm}
\usepackage{mathtools}
\usepackage{amsfonts}
\usepackage{comment}
\usepackage{bm}
\usepackage{ulem}
\newcommand{\be}{\begin{equation}}
\newcommand{\ee}{\end{equation}}
\newcommand{\bea}{\begin{eqnarray}}
\newcommand{\eea}{\end{eqnarray}}
\newcommand{\nn}{\nonumber}

\begin{document}
	\title{NANOGrav Hints on Planet-Mass Primordial Black Holes}

\author{\textsc{Guillem Dom\`enech$^{a}$}}
\author{\textsc{Shi Pi$^{b,c}$} }
\email{Corresponding author: shi.pi@itp.ac.cn} 

\affiliation{
$^{a}$\small{INFN Sezione di Padova, I-35131 Padova, Italy}\\
$^{b}$\small{CAS Key Laboratory of Theoretical Physics, Institute of Theoretical Physics, Chinese Academy of Sciences, P.O. Box 2735, Beijing 100190, China}\\
$^{c}$\small{Kavli Institute for the Physics and Mathematics of the Universe (WPI), The University of Tokyo Institutes for Advanced Study, The University of Tokyo, Kashiwa, Chiba 277-8583, Japan}
}

\begin{abstract}
Recently, the North American Nanohertz Observatory for Gravitational Waves (NANOGrav) claimed the detection of a stochastic common-spectrum process of the pulsar timing array (PTA) time residuals from their 12.5 year data, which might be the first detection of the stochastic background of gravitational waves (GWs). We show that the amplitude and the power index of such waves imply that they could be the secondary GWs induced by the peaked curvature perturbation with a dust-like post inflationary era with $-0.091\lesssim w\lesssim0.048$. Such stochastic background of GWs naturally predicts substantial existence of planet-mass primordial black holes (PBHs), which can be the lensing objects for the ultrashort-timescale microlensing events observed by the Optical Gravitational Lensing Experiment (OGLE).
\end{abstract}


	\maketitle

\section{Introduction}
After LIGO/VIRGO have detected gravitational waves (GWs) from mergers of  black holes and neutron stars~\cite{Abbott:2016blz,Abbott:2016nmj,Abbott:2017vtc,Abbott:2017gyy,Abbott:2017oio,TheLIGOScientific:2017qsa}, the next inspiring discovery may be the stochastic gravitational wave background (SGWB), which spans a large frequency range from $10^{-20}$ Hz to $10^{8}$ Hz. 
The pulsar timing array (PTA) can detect the low-freqency band of the SGWB down to $10^{-8}$ Hz, which is a good window for GWs from the early universe~\cite{Sazhin:1977tq,Detweiler:1979wn}. 
Recently, the North American Nanohertz Observatory for Gravitational Waves (NANOGrav) claimed that they have detected a stochastic common-spectrum process from the NANOGrav 12.5-yr data set~\cite{Arzoumanian:2020vkk}. The quadrupolar spatial correlation~\cite{Hellings:1983fr} as evidence for the (SGWB) is not found, which still awaits future detections with higher precision. However, SGWB as the source of such stochastic signals is worth considering at the current stage, which when written in the spectrum of its energy density, $\Omega_\text{GW}$, has an amplitude from $1.19\times10^{-9}$ to $4.51\times10^{-9}$ at 2-$\sigma$ confidence level (CL) at the fiducial frequency $f_\text{yr}=1~\text{yr}^{-1}$. The shape of the GW spectrum can be fit by power-law as $\Omega_\text{GW}\propto f^\beta$ with $-1.5<\beta<0.5$, if only the first five bins of the signal are taken into consideration, as the high frequency bins are probably from noise~\cite{Arzoumanian:2020vkk}.

The smallness of $\beta$, especially its compatibility with a scale-invariant GW spectrum, motivates theorists to consider this signal to be the SGWB from cosmic strings~\cite{Ellis:2020ena,Blasi:2020mfx,Buchmuller:2020lbh,Samanta:2020cdk}, the first order phase transitions \cite{Nakai:2020oit,Addazi:2020zcj,Neronov:2020qrl,Ratzinger:2020koh,Bian:2020bps,Li:2020cjj,1821775,1821800}, or induced GWs \cite{Vaskonen:2020lbd,DeLuca:2020agl,Bhaumik:2020dor,Kohri:2020qqd,Vagnozzi:2020gtf,Namba:2020kij,Sugiyama:2020roc,1821793}. In the last case, 
$\Omega_\text{GW}$ goes like $\propto f^3$ in the infrared, if the universe is dominated by radiation when the corresponding wavelength reenters the Hubble horizon \cite{Cai:2019cdl}. In the ultraviolet, the GW spectrum usually decays exponentially or as power-law, depending on the shape of the peak of the scalar perturbation, which is highly model-dependent~\cite{Pi:2020otn}. Therefore, the power index of $\Omega_\text{GW}$ can only fall in $-1.5<\beta<0.5$ in the near-peak frequency range, which put strong constraints on solar mass PBHs~\cite{Kohri:2020qqd}, unless the peak in the scalar power spectrum is broad enough~\cite{DeLuca:2020agl,Sugiyama:2020roc,Vaskonen:2020lbd}.

The infrared power index of $\Omega_\text{GW}\propto f^\beta$, however, depends sensitively on the equation of state (EoS) of the universe when the corresponding wavelength reenters the Hubble horizon. While the minimal cosmological model assumes the universe to be dominated by radiation from the end of inflation to the onset of matter domination, there is only solid evidence that the universe was dominated by radiation during big bang nucleosynthesis (BBN). 
Before that, there is little observational constraints, 
which affords fruitful phenomena for a prior stage of $w\neq1/3$. For a review, see \cite{Allahverdi:2020bys} and references therein.

The SGWB provides an opportunity to directly probe the expansion history of this primordial dark universe~\cite{Seto:2003kc,Boyle:2005se,Giovannini:2008tm,Guzzetti:2016mkm,Cui:2017ufi,Cui:2018rwi,Caprini:2018mtu,Escriva:2019phb,Cai:2019cdl,DEramo:2019tit,Hajkarim:2019nbx,Kapadia:2020pnr,Blasi:2020wpy,Domenech:2019quo,Domenech:2020kqm,Dalianis:2020cla}.
An EoS different from radiation, i.e. $w\neq1/3$, can be realized, e.g., by an adiabatic perfect fluid, or by a scalar field either oscillating in an arbitrary potential or rolling down an exponential potential \cite{Lucchin:1984yf}. For the induced GWs, the infrared power index $\beta$ will change at the reheating frequency which corresponds to the mode that reenters the horizon when the universe becomes radiation dominated.
If $w$ is small, the infrared shape of the GW spectrum can be flat enough to be consistent with the NANOGrav data, which also leaves a much smaller PBH mass than the one solar mass corresponding to the PTA frequency of $10^{-8}$ Hz. In this paper, we show that the recently reported NANOGrav signal is consistent with such a scenario, while its amplitude and power index predict a substantial amount of planet-mass PBHs, which is also implied by the ultrashort-timescale microlensing events recently observed by the Optical Gravitational Lensing Experiment (OGLE)~\cite{2017Natur.548..183M}.


\section{Induced GWs}
Gravitational waves induced by a scalar perturbation that peaks at $k_*$ are mainly 
generated when the $k_*$-mode reenters the horizon \cite{Matarrese:1992rp,Matarrese:1993zf,Matarrese:1997ay,Noh:2004bc,Carbone:2004iv,Nakamura:2004rm,Ananda:2006af,Baumann:2007zm,Osano:2006ew}.\footnote{This applies only for $w\leq 1/3$. When $w>1/3$ the scalar mode with $k=k_*$ keeps sourcing tensor modes.} After the generation, the evolution of tensor modes is essentially that of a massless free field. This means that after a tensor mode with wavenumber $k$ reenters the horizon, its energy density redshifts as radiation, i.e. $\rho_{\rm GW}\propto a^{-4}$. Thus, in a radiation dominated universe the spectrum of GW fractional energy density per logarithmic wavenumber interval (or GW spectrum for short), $\Omega_\text{GW}\equiv\rho_\text{total}^{-1}d\rho_\text{GW}/d\ln k$, does not redshift after generation. The interesting point is that any deviation from the standard radiation dominated universe is imprinted in a change of slope of the GW spectrum. The spectrum of induced GWs is then potentially compatible with the NANOGrav 12.5-yr result for certain expansion histories.

Considering that the expansion rate goes as $H^2\propto a^{3(1+w)}$ for a constant $w$, we find that a given comoving wavenumber $k$ is related to the scale factor at horizon crossing by $k\propto a_k^{-(1+3w)/2}$. Taking into account that GWs redshift as $a^{-4}$ after they reenter the horizon, we obtain $\Omega_{\rm GW}\propto k^{3}\left(a_k/a\right)^{1-3w}\propto k^{3-2\frac{1-3w}{1+3w}}\equiv k^\beta$\cite{Cai:2019cdl,Domenech:2020kqm,Hook:2020phx}, 
where $k^3$ comes from causality arguments for a localized source, i.e. a peak with finite time and spectral duration~\cite{Caprini:2009fx,Cai:2019cdl}\footnote{A detailed computation yields $\Omega_{\rm GW}\propto k^{3-2\left|({1-3w})/({1+3w})\right|}$ due to an extra superhorizon growth for $w>1/3$ \cite{Domenech:2020kqm}. We will not pursue the cases $w>1/3$ since the NANOGrav result would imply $w>1$. In the power-law scalar field model this corresponds to a negative potential which we regard as unphysical.}. 
It becomes $k^2$ in the near-infrared region of the induced GWs from a narrow peak~\cite{Cai:2019cdl,Pi:2020otn},
which gives
\begin{align}\label{eq:betaw}
\beta&=2-2\frac{1-3w}{1+3w},\quad\left(\Delta<k/k_*<1\right)
\end{align}
where $\Delta\lesssim1$ is the dimensionless width of the scalar perturbation power spectrum with a lognormal peak, 
$\mathcal{P_R}=(2\pi)^{-1/2}(\mathcal{A}_\mathcal{R}/\Delta)\exp\left[-\ln^2(k/k_*)/(2\Delta^2)\right]$. 
We see that for $w<1/3$ the spectral index of the GWs spectrum is less than in a radiation dominated universe due to a slower expansion rate. 

From now on, we consider for simplicity that the reheating temperature is lower than $130$ MeV given by  the lowest NANOGrav frequency bin, i.e. $2.4\times10^{-9}{\rm Hz}$, which is still allowed by BBN constraint of $T_{\rm rh}\gtrsim4~{\rm MeV}$~\cite{Kawasaki:1999na,Kawasaki:2000en,Hannestad:2004px,Hasegawa:2019jsa}. Such low-temperature reheating is well motivated by the decay of the light modulus fields, which are necessary ingredients of the low-energy realization of string theory~\cite{Kane:2015jia}. 
This gives
\begin{align}
f\approx 2.35\times 10^{-9}{\rm Hz}&\left(\frac{k}{k_{\rm rh}}\right)\left(\frac{T_{\rm rh}}{130{\rm MeV}}\right)\nonumber\\&\times
\left(\frac{g_*(T_{\rm rh})}{13.5}\right)^{1/2}\left(\frac{g_{*,s}(T_{\rm rh})}{14.25}\right)^{-1/3}.
\end{align}
Under this assumption, 
by using Eq.~\eqref{eq:betaw}, we find that the $1$-$\sigma$ constraints of the power index by NANOGrav, i.e. $0.5>\beta>-1.5$, can be converted to the range of 
the EoS parameter $w$ of 
\begin{align}
\label{wrangenarrow}
-0.091<w<0.048\,,
\end{align} 
for the near-infrared band of a narrow peak ($\Delta<k/k_*<1$), and $-0.128<w<-0.037$ for the other cases, i.e. far-infrared band of a narrow peak ($k/k_*<\Delta<1$) and infrared band of a broad peak ($\Delta\gtrsim1$).
\begin{figure}
\includegraphics[width=\columnwidth]{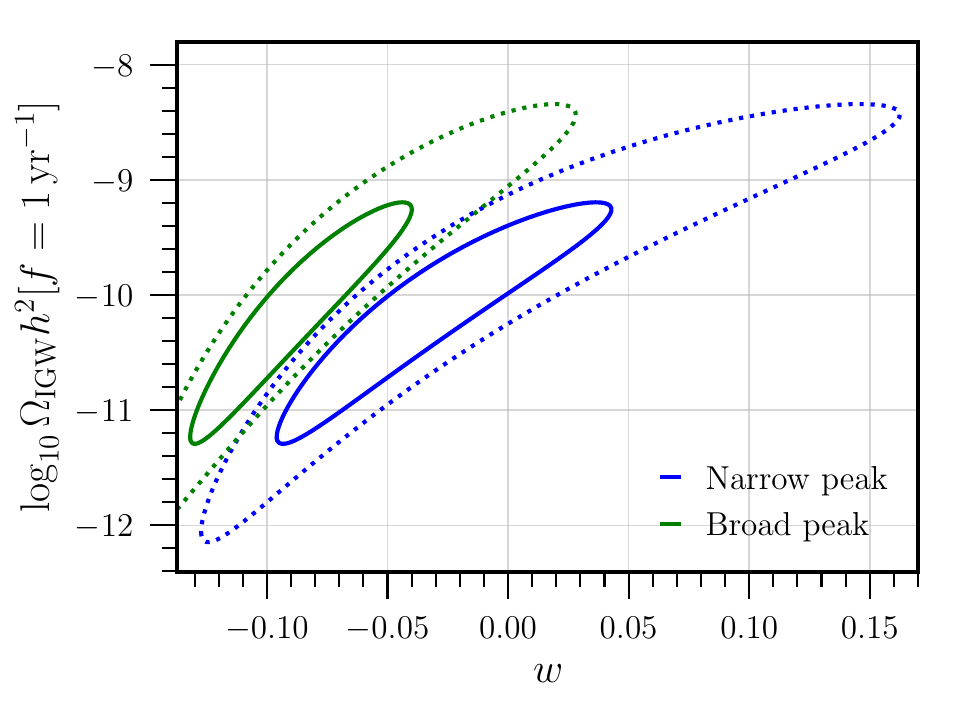}
\caption{$1$- and $2$-$\sigma$ contours of the posteriors for the amplitude of the GWs power spectrum at $f=1\,{\rm yr}^{-1}$ and the EoS parameter $w$, for a common-spectrum process for the five frequency power law of NANOGrav 12.5-yr results \cite{Arzoumanian:2020vkk}, respectively in solid and dotted lines. In blue and green we respectively show the implication for a narrow and a broad peak in the primordial scalar power spectrum. Note how the mean values for both cases falls in the region where $w<0$. We approximated the NANOGrav $1$ and $2$-$\sigma$ contours \cite{Arzoumanian:2020vkk} by ellispes.}\label{fig:w}
\end{figure}
 
We show the $1$- and $2$-$\sigma$ contours in Fig.~\ref{fig:w} extracted from the NANOGrav results. Notably, both mean values corresponds to a negative EoS parameter. In such a case, 
an adiabatic perfect fluid description is not appropriate as its negative sound speed, i.e. $c_s^2=w<0$, causes a pathological tachyonic instability. Therefore, we consider that the universe is dominated by a scalar field in an exponential potential \cite{Lucchin:1984yf}, which we dubbed as the $w$-dominated universe. It has the same background expansion as an adiabatic perfect fluid but differs at the perturbation level, which is well behaved with $c_s=1$ even when $w<0$. A remarkable property is that as $c_s=1$, the Jeans length is always higher than that of a perfect fluid, which makes the PBHs more difficult to form.

\begin{figure}
\includegraphics[width=\columnwidth]{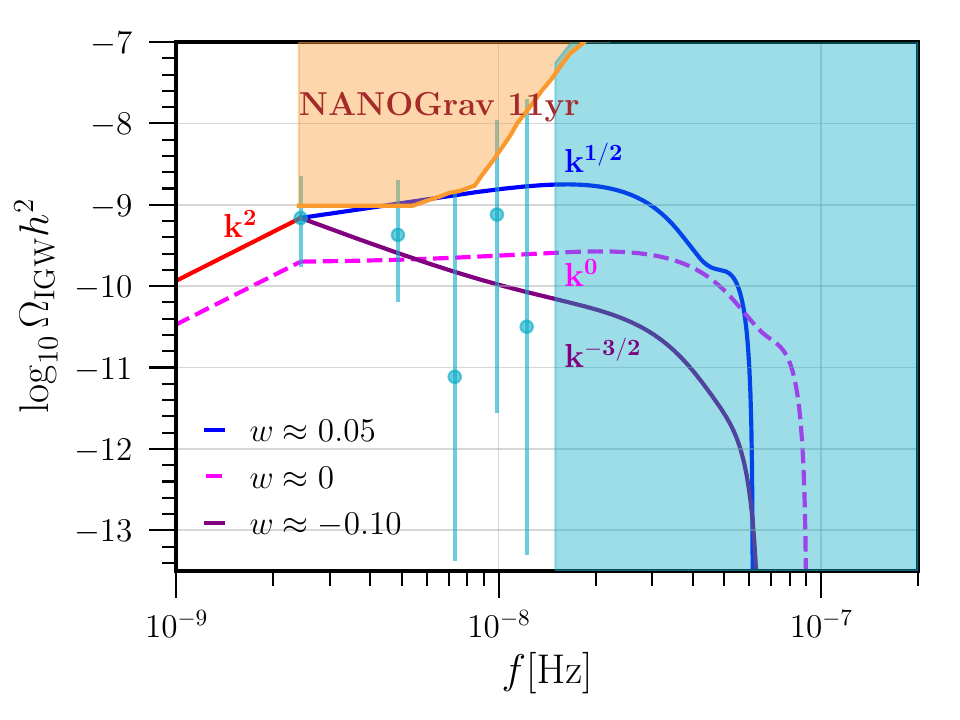}
\caption{GWs spectrum vs frequency. The solid blue and purple lines correspond to the scalar induced GWs in a $w$-dominated universe, respectively for $w\approx0.05\,,-0.1$. In the blue and purple solid lines we chose $k_{*}/k_{\rm rh}=10$ and respectively ${\cal A}_{\cal R}=0.14,0.26$. We also plot in magenta the characteristic value that fits the OGLE observation of PBH candidates with ${\cal A}_{\cal R}=0.14$, $k_{*}/k_{\rm rh}=14$ and $w=0$. In cyan we illustrate the NANOGrav 12.5-yr results for the residuals \cite{Arzoumanian:2020vkk} translated to the GW spectrum: the cyan dots represent the mean values of the residuals for the five frequency window, while the cyan shaded region illustrates the data which fits to white noise and is not used for the NANOGrav power-law fit. For comparison with previous results, we show in orange the sensitivity curve of the NANOGrav 11-yr results for a free spectrum \cite{Arzoumanian:2018saf}. \label{fig:Omega}}
\end{figure}

To estimate the amplitude of the induced GWs in a $w$-dominated universe we use the analytical results given in Ref.~\cite{Domenech:2020kqm}. For simplicity, we assume that the primordial scalar power spectrum has a narrow peak at $k=k_*$, which is described by a $\delta$-function as $\mathcal{P_R}=\mathcal{A_R}\delta(\ln(k/k_*))$.  We also assume the universe is reheated instantenously, when the $k_\text{rh}$-mode reenters the horizon. For a narrow peak, the spectrum of induced GWs near the scale of reheating $k_{\rm rh}$ and far enough from the peak scale $k_*$ takes the form of a broken power-law given by
\begin{align}\label{eq:OMbp}
&\Omega_{\rm GW}h^2(k\ll k_*)\approx\nonumber\\
&\Omega_{\rm GW,rh}h^2\times
\left\{
\begin{matrix}
&\displaystyle\hspace{-3em}\left(\frac{4k}{3k_{\rm rh}}\right)^{2} &(k\lesssim\frac34k_{\rm rh})\\
\\
&\displaystyle\left(\frac{4k}{3k_{\rm rh}}\right)^{2-2\frac{1-3w}{1+3w}} &(k\gtrsim \frac34k_{\rm rh})
\end{matrix}
\right.\,,
\end{align}
where
\begin{align}\nn
\Omega_{\rm GW,rh}h^2&\approx3.51\times10^{-7}\,\left(\frac{\Omega_{r,0}h^2}{4.18\times10^{-5}}\right)\left(\frac{g_{*,s}(T_{\rm rh})}{14.25}\right)^{-4/3}\\\label{eq:Orh}
&\times\left(\frac{g_{*}(T_{\rm rh})}{13.5}\right)\left(
\frac{9(1+w)(1+3w^2)}{4(1-3w)}
{\cal A}_{\cal R}\frac{k_{\rm rh}}{k_*}\right)^2\,.
\end{align}
The effective degrees of freedom $g_*$ and $g_{*s}$ must be evaluated at $T_{\rm rh}$, which is lower than $130~\text{MeV}$ as we stated before. An accurate position of the matching is numerically found to be $\sim 3k_{\rm rh}/4$ independent of $w$ \cite{Domenech:2020kqm}. Note that the amplitude of the GW spectrum is suppressed by a factor $(k_{\rm rh}/k_*)^2$, which can be easily understood as follows. Before the scalar mode with wavenumber $k_*$ enters the horizon, a tensor mode with wavenumber $k$ has a constant source leading to a growth proportional to $(k\tau)^2$. Once the scalar mode $k_*$ reenters the horizon at $\tau_*\sim1/k_*$, the source effectively shuts off, which yields the $\left(k/k_*\right)^2$ dependence. This implies that the further the peak scale is from the scale of reheating, the lower the amplitude of the scalar perturbation, if the amplitude of the GW spectrum at the reheating wavenumber is fixed. See Fig. \ref{fig:Omega}.

We can estimate the model parameters by using the $2$-$\sigma$ limits on the amplitude of the SGWB, which roughly translates to $\Omega_{\rm GW,rh}h^2\in[2.13, 8.08]\times10^{-10}$ at $f\approx2.4\times10^{-9}~{\rm Hz}$ \cite{Arzoumanian:2020vkk}. Using Eq.~\eqref{eq:Orh} we see that in order to reach the NANOGrav result, the product of the amplitude of the scalar peak and the separation of scales $k_{\rm rh}/k_*$ is a function of $w$
\begin{align}\label{eq:connexion}
	{\cal A}_{\cal R}\frac{k_{\rm rh}}{k_*}\approx(3.6\pm1.2)\times 10^{-2}\frac{4(1-3w)}{9(1+w)(1+3w^2)},
\end{align}
where we already took the best fit of $\Omega_{r,0}h^2$, $g_{*}$ and $g_{*s}$ in Eq.~\eqref{eq:Orh}. This is valid for $w<1/3$, which includes the parameter space of our interests shown in Eq.~\eqref{wrangenarrow}.

We see from Eq.~\eqref{eq:connexion} that there is a degeneracy of $\mathcal{A_R}$ and $k_{\rm rh}/k_*$, given that the amplitude of the induced GWs at $k_\text{rh}$ is fixed to be what NANOGrav has measured. Besides, the fact that the signal observed by NANOGrav spans at least for the first five bins in the range of $2.4\times10^{-9}~\text{Hz}\sim1.3\times10^{-8}$ Hz means that there is a lower bound for $k_*/k_\text{rh}$,
\be
\frac{k_*}{k_\text{rh}}>\frac{1.3\times10^{-8}}{2.4\times10^{-9}}\approx5.4,
\ee
which can be converted by \eqref{eq:connexion} to be a lower bound for $\mathcal{A_R}$, namely
\be\label{ARlower}
\mathcal{A_R}>4.0\times10^{-2}.
\ee
As we proceed to show in the next section, such a peak can generate substantial planet-mass PBHs, which, as implied by the recent detections of ultrashort-timescale microlensing events in OGLE data, consist $\sim2\%$ of the dark matter. 

\section{Primordial Black Holes}
As we have stated, when $-0.090<w<0.048$, the peak of the scalar power spectrum is to the ultraviolet of the peak of the induced GWs, and the GW spectrum between these two peaks has a power index that lies in $[-1.5,~0.5]$ to 1-$\sigma$ confidence level. These two peak frequencies as well as their amplitudes are connected by Eq.~\eqref{eq:connexion}, and the might-be detection of a relatively large GW spectrum implies that this frequency difference is not large, which can be further constrained by the PBH abundance. 


PBHs will form by gravitational collapse at horizon reentry for the Hubble patches where the density contrast $\delta\rho/\rho$ exceeds a threshold $\delta_c$~\cite{Zeldovich:1967lct,Hawking:1971ei,Carr:1974nx,Meszaros:1974tb,Carr:1975qj,Khlopov:1985jw}, which depends on the EoS parameter $w$ at the re-entry moment~\cite{Musco:2004ak,Musco:2008hv,Musco:2012au,Harada:2013epa}. As a conservative estimate we use the Carr's criterion in the uniform Hubble slice $\tilde\delta_c=c_s^2=1$~\cite{Carr:1975qj}, which when transferred to the comoving slice gives 
\begin{align}\label{dcb}
\delta_c=\frac{3(1+w)}{5+3w}\,.
\end{align}
The abundance of PBHs at their formation can be calculated by the Press-Schechter theory\footnote{See Refs.~\cite{Germani:2019zez,Escriva:2019phb} for recent progress within peak theory.}~\cite{Sasaki:2018dmp}
\be\label{betamono}
\beta(M_\mathrm{PBH})=\frac{\gamma}{2}\text{erfc}\left(\frac{\delta_c(w)}{\sqrt2\sigma(M_\text{PBH})}\right),
\ee
where $\gamma\approx0.2$ is the fraction of matter inside the Hubble horizon that collapses into PBH. $\sigma_\text{PBH}(M_\text{PBH})$ is the variance of the density perturbation smoothed on the mass scale of $M_\text{PBH}$,
\be
\sigma^2=\left(\frac{2(1 + w)}{5 + 3 w}\right)^2\int d\ln k~W^2(kR)(kR)^4\mathcal{P}_\mathcal{R}(k),
\ee
with $W(kR)$ the window function to smooth the perturbation on the comoving scale $R=2GM_\mathrm{PBH}/a_\text{form}$. We take $W(kR)=\exp(-k^2R^2)$ in this paper, 
though different choices of window functions will leave significant uncertainties in the PBH abundance \cite{Tokeshi:2020tjq}. 

\begin{figure}
\includegraphics[width=\columnwidth]{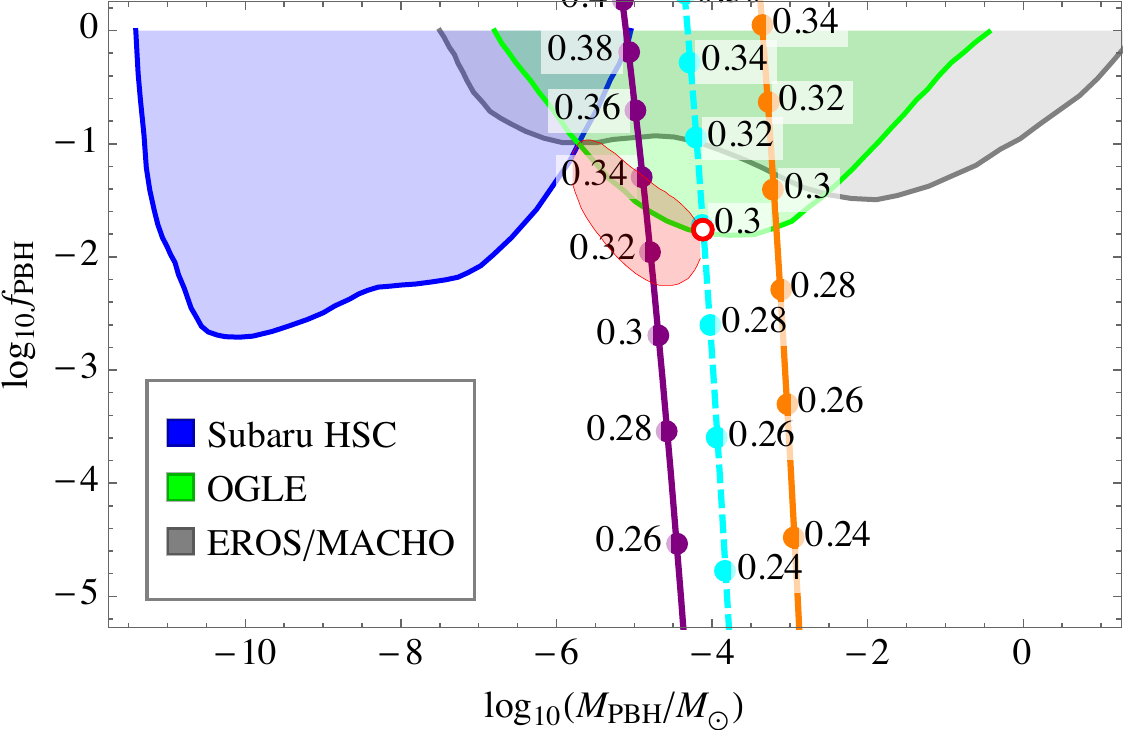}
\caption{The PBH abundance as a function of its mass. The blue, green, and gray shaded areas are excluded by current microlensing experiments of Subaru HSC~\cite{Niikura:2017zjd}, OGLE~\cite{Niikura:2019kqi}, and EROS/MACHO~\cite{Tisserand:2006zx}, respectively. The region allowed by 1-$\sigma$ power index and 2-$\sigma$  characteristic strain from NANOGrav result is between the two solid lines, which are $w=-0.091$, $h_c=1.37\times10^{-15}$ (left, purple) and $w=0.048$, $h_c=2.67\times10^{-15}$ (right, orange), respectively. The cyan dashed line is the 2-$\sigma$ lower bound of $h_c$ with $w=0$. Some values of $\mathcal{A_R}$ are labeled on the lines. The red contour is the 95\% CL PBH abundance, if the six planet-mass lensing objects found in OGLE~\cite{2017Natur.548..183M} are PBHs~\cite{Niikura:2019kqi}. The small circle denotes a typical value of $f_{\text{PBH}}=0.0174$ at $M_{\text{PBH}}=7.60\times10^{-5}M_\odot$.}
\label{fig:f(M)}
\end{figure}

In a general $w$-dominated universe, the masses of the PBHs are related to the comoving scale by
\begin{align}\label{mpbh}
\frac{M_{\rm PBH}}{M_\odot}&\approx  1.58\frac{\gamma}{0.2}\left(\frac{k_{\rm rh}}{k_*}\right)^{\frac{3(1+w)}{1+3w}}
\left(\frac{13.5}{g_{*}(T_{\rm rh})}\right)^{\frac12}\left(\frac{130{\rm MeV}}{T_{\rm rh}}\right)^{2}
\end{align}
The mass function of the PBHs, i.e. PBH energy fraction with respect to cold dark matter, is given by
\begin{align}\label{fpbh}
f_{\rm PBH}\equiv\frac{\Omega_\text{PBH}}{\Omega_\text{CDM}}&\approx 1.96\,\beta\times 10^{12}\left(\frac{k_*}{k_{\rm rh}}\right)^{\frac{6w}{1+3w}}\left(\frac{T_{\rm rh}}{130{\rm MeV}}\right)\nonumber\\&\times\left(\frac{g_{*,s}(T_{\rm rh})}{14.25}\right)^{-1}\left(\frac{g_{*}(T_{\rm rh})}{13.5}\right)\,,
\end{align}
where $\beta$ is given by \eqref{betamono}. A very sharp peak in the primordial scalar power spectrum leads to a monochromatic mass function. 
We first take $T_\text{rh}\approx130$ MeV with the corresponding values of $g_*$ and $g_{*s}$, then use \eqref{eq:connexion} and \eqref{dcb} to replace $k_\text{rh}/k_*$ and $\delta_c$ with $\mathcal{A_R}$ and $w$. We see that both $f_\text{PBH}$ and $M_\text{PBH}$ are functions of $\mathcal{A_R}$ and $w$, while $w$ is bounded by \eqref{wrangenarrow}. The lower bound of $\mathcal{A_R}$ is given by \eqref{ARlower}, while its upper bound is given by the observational constraints on the PBH abundance. 
Putting together \eqref{mpbh} and \eqref{fpbh}, we can find the function of $f_\text{PBH}(M_\text{PBH})$, which is drawn in Fig.\ref{fig:f(M)} with different values of $\mathcal{A_R}$ and $w$, together with the current microlensing constraints on the PBH abundance. A very interesting fact is that the mass window of the PBHs allowed by the NANOGrav result, $10^{-6}M_\odot$ to $10^{-2}M_\odot$ (planet mass), is consistent with the implication of the ultrashort-timescale microlensing events recently observed by OGLE~\cite{2017Natur.548..183M,Niikura:2019kqi}, 
if
\be\label{range:w}
-0.091\lesssim w\lesssim0.
\ee
Especially, $w=0$ is critically allowed in the 2-$\sigma$ region of the PBH abundance. A typical parameter choice is
\begin{align}
\mathcal{A_R}=0.11, \quad w=0,
\end{align}
which gives $f_\text{PBH}=1.74\times10^{-2}$ at $M_\text{PBH}=7.60\times10^{-5}M_\odot$, marked as a small red circle in Fig.\ref{fig:f(M)}. 

Therefore, when there is a dust-like stage before radiation dominated era, both the common-spectrum process of time residuals (as SGWB) observed by NANOGrav and the ultrashort-timescale microlensing events (as planet-mass PBHs) observed by OGLE originates from the same peaked scalar perturbation. A low reheating temperature smaller than $130$ MeV and a nearly dust-like stage with $-0.091\lesssim w\lesssim0$ between the reheating moment and the peak-reentry moment are crucial to make this happen. 
We would like to mention that both of the observations on the GWs and PBHs are not clear evidence. For NANOGrav signals, no Hellings-Downs correlation~\cite{Hellings:1983fr} is found, while for OGLE results the contamination of unbounded (wide-orbit or free-floating) planets in the galactic disk must be removed. 
Both of these can be cleared in the future experiments. 
For nanoHertz SGWB, FAST~\cite{Hobbs:2014tqa} and SKA~\cite{Janssen:2014dka} can increase the sensitivity for  3 to 5 orders. The microlensing search for  Andromeda Galaxy by Subaru HSC can greatly avoid the unbounded planets as they mainly exist in the galactic disk~\cite{Niikura:2019kqi,Kusenko:2020pcg}. 
Also, a PBH of $10^{-5}~M_\odot$ captured by the solar system as the Planet 9 can cause the orbital anomalies of the trans-Neptunian objects~\cite{Scholtz:2019csj}, whose surrounding minihalo~\cite{Siraj:2020upy}, Hawking radiation~\cite{Arbey:2020urq}, or gravitational field~\cite{Witten:2020ifl} could be probed directly. All of these experiments could verify or falsify the prediction of our scenario in the near future.

\section{Conclusion}
The fact that the power index of the GW spectrum fit from first five bins of the time residuals observed by NANOGrav is small 
motivates us to consider that NANOGrav has observed induced GWs from a (close to) dust-like stage. 
If such a stage is driven by a scalar field, this scenario predicts planet-mass PBHs whose masses are from $10^{-6}~M_\odot$ to $10^{-2}~M_\odot$, depending on the value of 
of the effective EoS parameter $w$. 
We find that for $-0.091\lesssim w\lesssim0$ the predicted PBH mass and abundance can also explain the 
recently detected ultrashort-timescale microlensing events from OGLE data. This prediction that both the stochastic GWs at around $10^{-9}$ Hz and the planet-mass PBHs are from the same origin is very intriguing, which can be further explored by the future experiments of PTA, microlensing, and direct search of the Planet 9. 

It should be noted that the PBH abundance is exponentially sensitive to the precise value of the density threshold. Thus, our result should be taken as a rough estimate. Numerical calculations of PBH formation in a scalar field domination are needed for a determination of $\delta_c$ and $\gamma$ with higher precision, especially for the $w$-dominated universe we considered. Also, our conclusion cannot be directly extrapolated to an adiabatic perfect fluid, which enhances the PBH formation when $w$ decreases, especially when the EoS is soften in the thermal history of the universe~\cite{Byrnes:2018clq,Carr:2019kxo}. In a matter-dominated universe similar to our case, the PBH formation is much enhanced~\cite{Khlopov:1980mg,Harada:2016mhb,Harada:2017fjm}, which is equivalent to suppress the amplitude of the curvature perturbation spectrum for a fixed PBH abundance and shift the PBH mass to higher regions. 

In this paper, we focused on a $\delta$-function peak in the curvature perturbation spectrum, of which the result is the same as that of the lognormal peak case
$\mathcal{P_R}=(2\pi)^{-1/2}(\mathcal{A}_\mathcal{R}/\Delta)\exp\left[-\ln^2(k/k_*)/(2\Delta^2)\right]$
with a narrow width $\Delta<k_\text{rh}/k_*\lesssim0.19$~\cite{Pi:2020otn}. This is because in this narrow-peak case, the GW spectrum in the PTA frequency band remains unaffected, while the far infrared power index at $f\ll f_*\Delta<f_\mathrm{rh}\lesssim2.4\times10^{-9}$ is increased by 1, which however is beyond the current observational frequency band of PTA. As the width becomes larger, i.e. 
$0.19\lesssim\Delta\lesssim0.4$, the knee frequency $f_*\Delta$ where the power index changes lies in $[f_\text{rh},f_*]$ and becomes observable \cite{Domenech:2020kqm}, which needs a new fit with a broken-power-law spectrum. A broader width of $\Delta\gtrsim0.4$ increases the entire power for $f<f_*$ by 1, which gives a different allowed region of $-0.014<w<-0.04$ (See Fig.\ref{fig:w}). In the broad-peak case the PBH formation and its observation should also be reexamined~\cite{Carr:2017jsz}. We leave this issue for future work. 


\textit{Acknowledgement}~~
We thank Sabino Matarrese and Misao Sasaki for useful discussions and comments. 
This work was supported by the National Key Research and Development Program of China (Grant No. 2020YFC2201502).
G.D. was partially supported by the European Union’s Horizon 2020 research and innovation programme under the Marie Sk{\l}odowska-Curie grant agreement No 754496. 
S.P. was supported by the Key Research Program of the Chinese Academy of
Sciences Grant NO. XDPB15, by the CAS Project for Young Scientists in Basic Research YSBR-006,
by Project 12047503 of the National Natural Science Foundation of China, by JSPS Grant-in-Aid for Early-Career Scientists No. JP20K14461, and by the World Premier International Research Center Initiative (WPI Initiative), MEXT, Japan.

\bibliography{biblio.bib}

\end{document}